# The origin of *s*-process isotope heterogeneity in the solar protoplanetary disk

Mattias Ek[1,2*], Alison C. Hunt[1], Maria Lugaro[3,4], Maria Schönbächler[1]

[1] Institute for Geochemistry and Petrology, ETH Zürich, Clausiusstrasse 25, 8092 Zürich, Switzerland.

[2] Bristol Isotope Group, School of Earth Sciences, University of Bristol, Wills Memorial Building, Queen's Road, Bristol BS8 1RJ, United Kingdom.

[3] Konkoly Observatory, Research Centre for Astronomy and Earth Sciences, MTA Centre for Excellence, H-1121 Budapest, Hungary.

[4] Monash Centre for Astrophysics, School of Physics and Astronomy, Monash University, VIC 3800, Australia.

*e-mail: mattias.ek@bristol.ac.uk

**Rocky asteroids and planets display nucleosynthetic isotope variations that are attributed to the heterogeneous distribution of stardust from different stellar sources in the solar protoplanetary disk. Here we report new high precision palladium isotope data for six iron meteorite groups, which display smaller nucleosynthetic isotope variations than the more refractory neighbouring elements. Based on this observation we present a new model in which thermal destruction of interstellar medium dust results in an enrichment of *s*-process dominated stardust in regions closer to the Sun. We propose that stardust is depleted in volatile elements due to incomplete condensation of these elements into dust around asymptotic giant branch (AGB) stars. This led to the smaller nucleosynthetic variations for Pd reported here and the lack of such variations for more volatile elements. The smaller magnitude variations measured in heavier refractory elements suggest that material from high-metallicity AGB stars dominated stardust in the Solar System. These stars produce less heavy *s*-process elements (Z ≥ 56) compared to the bulk Solar System composition.**

The protoplanetary disk from which our Solar System formed incorporated dust that was inherited from the collapsing molecular cloud. A few per cent of this dust formed around stars with active nucleosynthesis and retained the extreme isotopic fingerprint of its formation environment[1,2]. This dust, which is isotopically anomalous compared to Solar System compositions, is here termed stardust. However, it is often also referred to as presolar grains when found in meteorites. Most stardust in primitive meteorites originates from asymptotic giant branch (AGB) stars, the site of *s*-process nucleosynthesis, with only small contributions from supernovae environments[3]. The majority of the dust in the solar protoplanetary disk grew in the interstellar medium (ISM) from a well-mixed gas phase as mantles on pre-existing nuclei. These mantles likely inherited the composition of the local ISM, i.e., a near solar isotopic composition[1].

Nucleosynthetic isotope variations, relative to Earth, are well established for a range of elements in bulk meteorites[4]. These variations mostly reflect the heterogeneous distribution of isotopically distinct dust in the protoplanetary disk. It is generally thought that this heterogeneity was established, at least in part, due to processes occurring in the protoplanetary disk itself. Physical sorting of grains, either by mineralogical type[5] or size[6], and selective destruction of stardust by thermal processing in the protoplanetary disk[7-9] or by aqueous alteration on parent bodies[10], have all been proposed as possible mechanisms to generate isotope heterogeneity. While these individual processes can explain nucleosynthetic variations for specific elements, it is debated whether a unifying explanation for all elemental trends exists[11]. Therefore, considerable uncertainty remains as to which processes were important for dust processing in the protoplanetary disk.

Nucleosynthetic variations in bulk meteorites are mainly limited to refractory elements[4]. The neighbouring refractory elements Zr, Mo and Ru display well-defined and linearly correlated nucleosynthetic variations[9,11-14]. Each meteorite group shows a distinct *s*-process deficit in each of these elements, relative to Earth, which increases with the inferred formation distance of the meteorite parent body from the Sun[9,14,15]. This can be accounted for by a heterogeneous distribution of isotopes of *s*-process origin in the Solar System. Moreover, no nucleosynthetic variations are reported for more volatile elements in the same mass region, e.g., Cd[16,17] and Te[18], suggesting that the elemental condensation temperature plays a role in preserving nucleosynthetic variations. Palladium falls between Cd, Te and the refractory elements Zr, Mo and Ru in volatility, making it ideal for testing the link between volatility and the origin of the nucleosynthetic *s*-process variations. Earlier isotope data limited to the IVB iron meteorite group indicate that nucleosynthetic Pd offsets are smaller than predicted from Ru and Mo *s*-process deficits[19]. Iron meteorites are enriched in siderophile elements such as Ru, Pd and Mo and show distinct nucleosynthetic variations for Mo[14] and Ru[13,15]. They are thus ideal to constrain nucleosynthetic Pd isotope variations and their link to volatility.

**Results**

We determined Pd isotope compositions for 24 samples from the IAB, IIAB, IID, IIIAB, IVA and IVB groups (Supplementary Table 1; Methods). All data are presented in ε-notation, where ε is the deviation of the sample isotope ratio from the terrestrial standard NIST SRM 3138 in parts per 10,000 (see Methods). Several iron meteorite groups show isotopic variations between their members, most notably the IID and IVB irons (Extended Data Figure 1). These variations are not related to nuclear field shift effects (Extended Data Figure 2; Methods), however, they correlate with Pt isotope ratios, which are affected by exposure to galactic cosmic rays and provide an established cosmic ray exposure dosimeter (Extended Data Figure 3; Methods). After correction for cosmic ray irradiation, each group except the IAB irons exhibits a nucleosynthetic composition distinct from the terrestrial standard (Figure 1; Table 1). The negative $\varepsilon^{104}$Pd and $\varepsilon^{106}$Pd with concomitant positive $\varepsilon^{110}$Pd values indicate an *s*-process deficit in meteorites relative to Earth based on isotope yields of *s*-process models[20] (Figure 1). When plotting Mo, Ru and Pd isotope data together (Figure 2) we find a linear relationship that traces the processes that generated the nucleosynthetic heterogeneities in planetary bodies.

Table 1 | The nucleosynthetic Pd isotope composition of iron meteorite groups.

| Group | $\varepsilon^{102}$Pd | $\varepsilon^{104}$Pd | $\varepsilon^{106}$Pd | $\varepsilon^{110}$Pd |
|---|---|---|---|---|
| IAB | 0.20 ± 0.26* | 0.01 ± 0.12* | -0.01 ± 0.02* | -0.05 ± 0.05* |
| IIAB | 0.46 ± 0.60* | -0.12 ± 0.15* | -0.05 ± 0.07* | 0.05 ± 0.16* |
| IID | 0.20 ± 0.38* | -0.56 ± 0.12* | -0.08 ± 0.06* | 0.32 ± 0.12* |
| IIIAB | -0.05 ± 0.97* | -0.29 ± 0.26* | -0.01 ± 0.09* | 0.22 ± 0.19* |
| IVA | 0.11 ± 0.37* | -0.13 ± 0.07* | -0.02 ± 0.04* | 0.14 ± 0.09* |
| IVB | -0.52 ± 0.41* | -0.66 ± 0.22† | -0.09 ± 0.06† | 0.39 ± 0.13† |

*Average of unexposed samples (Supplementary Table 2). Uncertainties represent 2 standard errors of the mean for all analyses from unexposed samples calculated using the homoscedastic method (see Methods).

†Calculated via regression against $\varepsilon^{196}$Pt (Supplementary Table 2). Uncertainties represent the 2 standard deviation of the $\varepsilon^i$Pd intercept.

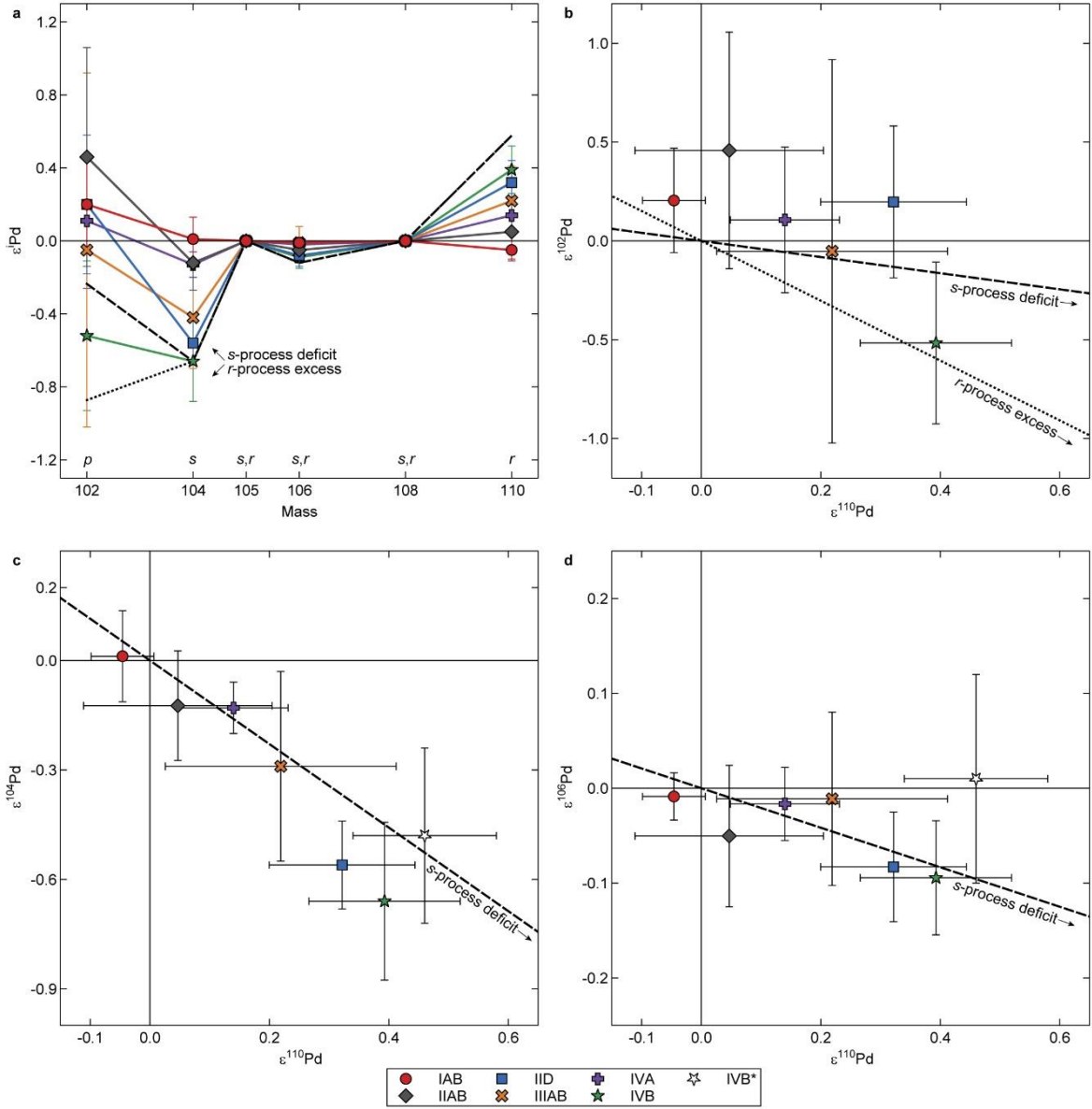

**Figure 1 | The nucleosynthetic Pd isotope composition of iron meteorites.** (**a**) The nucleosynthetic composition of the IAB, IIAB, IID, IVA, and IVB groups (Table 1). Graphs are shown for (**b**) $\varepsilon^{102}$Pd, (**c**) $\varepsilon^{104}$Pd and (**d**) $\varepsilon^{106}$Pd versus $\varepsilon^{110}$Pd normalised to $^{108}$Pd/$^{105}$Pd for the analysed meteorite groups after correction for cosmic ray exposure. Panel (**c**) and (**d**) also include the Pd data for the IVB group from Ref. 19 (denoted with a *). The primary nucleosynthetic source of each is isotope (*p*-, *s*-, or *r*-process) is stated above the mass in panel (**a**). Long dashed lines depict an *s*-process mixing line, while short dashed lines reflect *r*-process mixing lines calculated using Ref. 20. All data agree well with the trend predicted by the *s*-process mixing lines. The results favour an *s*-process deficit in iron meteorites relative to the Earth over an *r*-process excess, based on $\varepsilon^{102}$Pd. Our value for the IVB group overlaps within uncertainty with that reported by Ref. 19. The origin denotes the composition of the terrestrial isotope standard NIST SRM 3138. Uncertainties on data points reflect either the 2 standard error of the mean or the 2 standard deviation of the $\varepsilon^{i}$Pd intercept of a regression against $\varepsilon^{196}$Pt (see Table 1).

## Mixing interstellar medium grown dust with stardust

The linear correlation observed for samples from different bodies in our Solar System (Figure 2) provides evidence for mixing between two isotopically distinct reservoirs, with at least one clearly distinct from the bulk Solar System composition. Here, we propose that these two reservoirs are stardust (isotopically anomalous) and ISM dust whose isotopic composition closely resembles average Solar System composition. We argue that this stardust fraction mainly carries an *s*-process composition, because stardust with solar *p*- and *r*-process composition has not been unequivocally identified in meteorites. While some stardust grains show enrichments in *p*- and/or *r*-process isotopes, e.g. SiC[21,22] and nanodiamonds[23,24], their composition does not mirror the Solar System *r*-process isotopic component.

In addition, models of the weak *r*-process in core-collapse supernova neutrino winds do not predict isotopic abundances close to those of the Solar System *r*-process component[25]. Other models for the weak *r*-process, such as electron capture supernovae, favour the production of non-neutron-rich isotopes over classical *r*-process neutron-rich isotopes (e.g. $^{90}$Zr over $^{96}$Zr)[26], but these signatures are not observed in stardust either. Furthermore, in order to mimic the *s*-process pattern identified in our data, stardust needs to contain the same proportion of *p*- and *r*-process material as the bulk Solar System. This is an unrealistic assumption given that the *p*- process is believed to occur in core-collapse supernovae [27], while the *r*-process most likely occurs in neutron star mergers[28,29], two distinct stellar environments. A small enrichment of supernovae derived material in the outer Solar System was proposed to account for the isotope dichotomy between carbonaceous (CC) and non-carbonaceous meteorites (NC)[30-32]. If correct, such an enrichment can only constitute a minor addition to the overall stardust and/or ISM dust fractions (Extended Data Figure 4). Hence, contributions from *p*- and *r*-process material likely represent only a minor fraction of the isotopically anomalous material that makes up stardust.

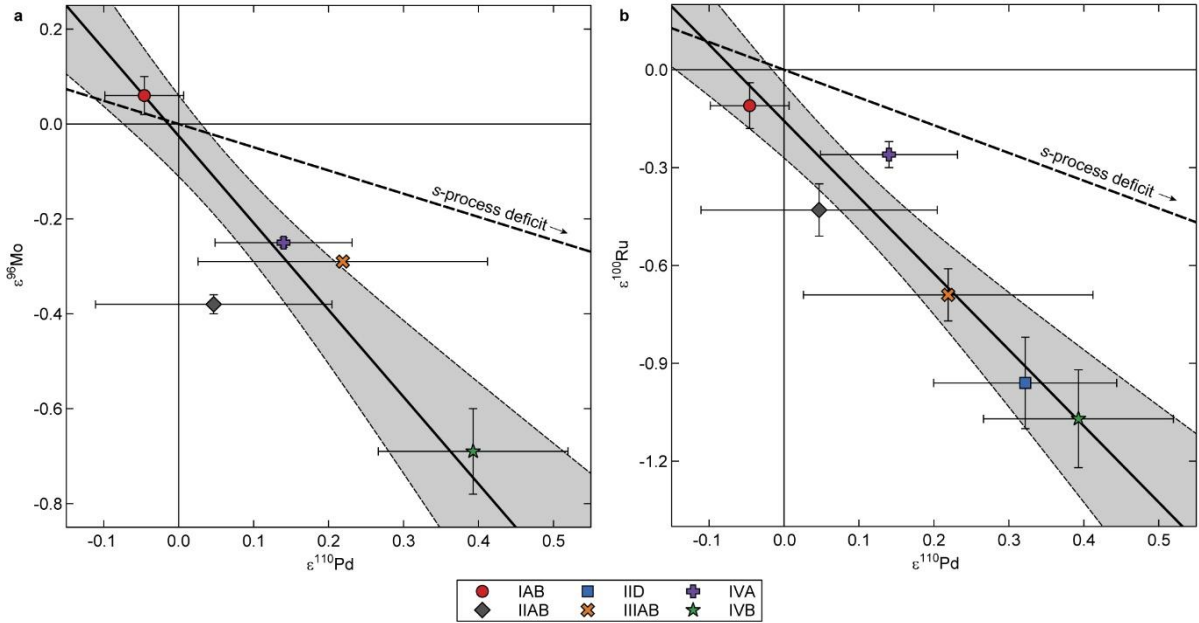

**Figure 2 | Correlation of $\varepsilon^{96}$Mo and $\varepsilon^{100}$Ru versus $\varepsilon^{110}$Pd for iron meteorite groups.** The solid line is the best-fit line for (**a**) $\varepsilon^{110}$Pd–$\varepsilon^{96}$Mo (slope = -1.83 ± 0.58 (2 σ), r$^2$ = 1) and (**b**) $\varepsilon^{110}$Pd–$\varepsilon^{100}$Ru (slope = -2.35 ± 0.67 (2 σ), r$^2$ = 1). Dashed lines represent the predicted mixing line calculated between the isotopic composition of Earth and an *s*-process component defined by Ref. 20 for (**a**) $\varepsilon^{110}$Pd–$\varepsilon^{96}$Mo (slope = -0.49) and (**b**) $\varepsilon^{110}$Pd–$\varepsilon^{100}$Ru (slope = -0.85). Palladium data are internally normalised to $^{108}$Pd/$^{105}$Pd. Molybdenum data (internally normalised to $^{97}$Mo/$^{95}$Mo) are from Ref. 11 and Ru data (internally normalised to $^{99}$Ru/$^{101}$Ru) are from Ref. 13,15. Uncertainties on Pd data points the same as in Figure 1. Uncertainties for $\varepsilon^{96}$Mo and $\varepsilon^{100}$Ru are given as the 2 standard error of the mean and the 95% confidence interval, respectively.

We therefore argue that the stardust reservoir, at first order, features an *s*-process composition and predominantly consists of material from AGB stars that became C-rich via recurrent mixing between the core and envelope (i.e., mass range 1.5 to 4 $M_\odot$). These stars are the source of the vast majority of presolar SiC grains identified in meteorites. The solar *s*-process component estimated from the modelled yields of such stars[20,33] reproduces the nucleosynthetic variations observed in Pd (Figure 1) and Zr[9], Mo[11,14] and Ru[13,15] isotopes well. AGB stars with other masses remain O-rich and are instead the source of the vast majority of oxide and silicate stardust[34]. This dust does not contain observable *s*-process elements and is therefore unlikely to significantly contribute elements heavier than Fe to stardust. Stardust contributions from the Light Element Primary Process (LEPP) sources[35] and spinstars[36] are also unlikely, because they existed mainly in the early Universe. Given that the mean residence time for stardust in the ISM is a few hundred million years[37], it is improbable that stardust from such environments survived until the formation of the Solar System. Any *s*-process nuclei produced in these environments are therefore part of the homogenised ISM dust fraction.

**Thermal processing of dust**

In the context of our model (Figure 3), the enrichment of *s*-process material in the inner Solar System was achieved through the destruction of ISM dust in this region by thermal processing such as photoevaporation induced by the radiation of the young Sun. Asteroids and planets forming in the inner region inherited this enrichment. Significant processing of silicates could occur as material accretes onto the disk[38] or during FU Orionis outbursts of the young Sun[39]. In line with this, previous studies proposed silicates as the likely phase being destroyed during thermal processing[7,9,31]. Another attractive mechanism is the removal of ISM-grown organic-rich icy mantles that are presumed to surround refractory dust grains[40]. Refractory elements can be implanted into such mantles by supernovae shockwaves in the ISM[41]. These volatile mantles will react to photoevaporation in the disk more rapidly than silicates and thus provide a wider temperature range in which thermal processing takes place while preserving the refractory stardust. Such a mechanism was previously suggested to explain volatile depletion in the Solar System[42]. Evidence for thermal processing of primitive Solar System materials is provided by noble gases and the abundances of SiC, presolar diamond and insoluble organic matter (IOM) in chondrites. Relative to CI chondrites, whose composition most closely matches that of the sun, many chondrite groups contain at least one dust component that has been processed at temperatures up to ~700 K[8,43]. This may, at least partly, reflect thermal processing in the solar protoplanetary disk[8]. Additionally, differences in the abundance of IOM and crystalline silicates between ISM material, primitive Solar System material such as comets and interplanetary dust particles, and CI chondrites suggests that a significant fraction of dust underwent thermal processing at temperatures of ~1000 K or higher in the disk[44]. Later mixing between thermally processed and relatively pristine material can account for the abundances of presolar components with different thermal susceptibilities observed in meteorites[44].

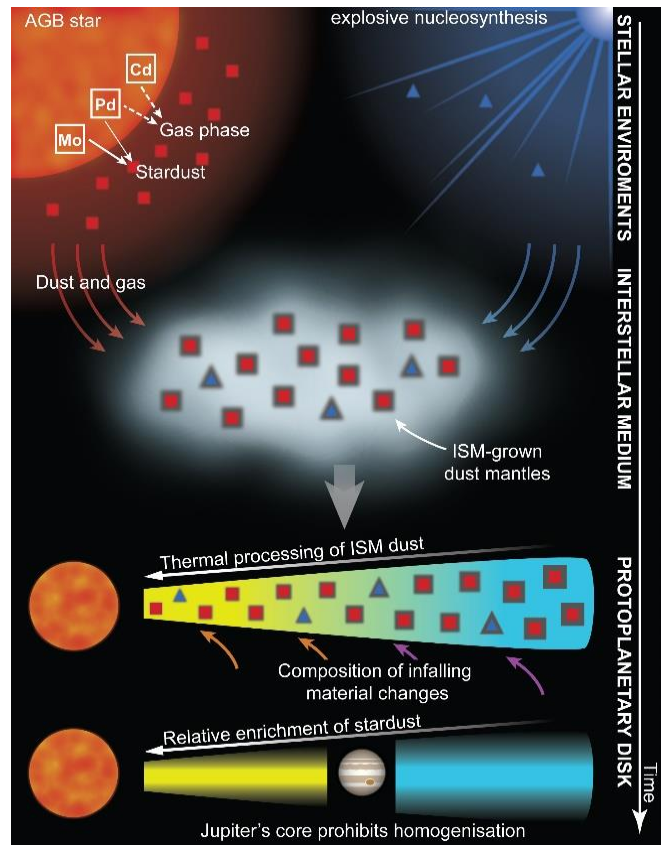

**Figure 3 | Cartoon illustrating dust formation and evolution as proposed in our model (not to scale).** Stardust forms in different stellar environments with ongoing nucleosynthesis and retains the unique isotope composition of its formation site. The trace element composition of AGB stardust depends on the condensation temperature of individual elements and is depleted in volatile elements compared to the bulk Solar System *s*-process component. Stardust in the Solar System mainly originates from low mass AGB stars that produce grains with an *s*-process isotope composition (red squares). In particular, Solar System stardust is sourced from high metallicity AGB stars that produce less heavy elements ($Z \geq 56$) compared to the average Solar System composition. Dust from other stellar environments (blue triangles), e.g. supernovae, contribute only a minor fraction to stardust. In dense molecular clouds, dust condenses from the homogenised gas phase onto pre-existing grains forming ISM dust/ice mantles around stardust (dark grey). Due to gravitational instabilities a part of the molecular cloud collapses and forms a protoplanetary disk surrounding a proto-star. Thermal gradients in the protoplanetary disk preferentially destroy ISM mantles in regions closer to the Sun (yellow regions), which results in a relative enrichment of stardust, predominantly carrying an *s*-process composition in the lighter elements ($Z<56$). It is possible that the composition of infalling material also changed with time and/or thermal processing[30,31,55]. Complete homogenisation between the inner and outer Solar System is blocked by the formation of Jupiter's core[55,56], which leads to two compositionally slightly different reservoirs.

We estimated the amount of ISM dust that was removed to recreate the 0.02 % *s*-process material excess in the Earth compared to carbonaceous chondrites[4]. About 0.37 % of the initial dust mass was removed in the form of ISM dust, assuming that stardust has the same isotope composition as the bulk Solar System *s*-process component, that stardust accounts for 3 % of the dust mass in the Solar System[2] and that the *s*-process accounts for 57.5 % of the Solar System mass for elements heavier than Fe[20] (see Methods for equations). This can be compared to the scenario where *p*- and *r*-process enriched stardust is removed (analogous to Equation 6 in Methods). This requires only 0.008 % mass loss of the initial dust (~44 times less compared to removing ISM dust). However, as discussed above, we discard this latter option because it is improbable that different *p*- and *r*-process sources would produce dust with similar thermal properties and in the same proportion as the solar *p*- and *r*-process component.

**Incomplete condensation of elements around AGB stars**

Noteworthy, the *s*-process deficits in Pd are smaller than predicted from the *s*-process depletion of Mo and Ru isotopes in the same meteorite groups (Figure 2), in agreement with a previous study of IVB irons[19]. This implies that the relative abundance of Pd, compared to Zr, Mo and Ru, in stardust is 3-4 times lower than in the bulk *s*-process composition of the Solar System. Furthermore, Cd and Te concentrations must be even more depleted in stardust because of the absence of nucleosynthetic variations in these elements in bulk meteorites[16,18]. Changes in stellar parameters including initial mass and metallicity do not affect the relative yields of Zr, Mo, Ru, Pd and Cd (Figure 4; Extended Data Figure 5) and can therefore not explain the reduced abundance of Pd and Cd in stardust. Likewise, enhanced rates of the $^{22}$Ne neutron source during AGB nucleosynthesis are unable to reduce the Pd yields relative to the neighbouring elements[19]. Therefore, the Pd depletion in stardust is not a consequence of stellar yields. We propose that it reflects incomplete condensation of elements into stardust around AGB stars as a function of the elemental condensation temperature (Figure 3). Incomplete condensation refers to the scenario where only a fraction of the elements in the gas phase condense into dust, while the gas phase is lost. The typical temperature of dust forming regions around AGB stars is ~1000 K[37], and hence a fraction of the elements with lower condensation temperature will likely be retained in the gas phase. Astronomical observations identified a distinct correlation between elemental abundances in the ISM gas phase and elemental condensation temperature[45], indicating that more refractory elements are preferentially incorporated into dust. Evidence that the trace element composition of stardust depends on the elemental condensation temperature is provided by presolar SiC grains that have been shown to favourably incorporate refractory elements[46-48]. Additionally, it has been inferred that shock processing of dust in the ISM increases the concentration of volatile elements in the gas phase[49]. This suggests that dust enriched in more volatile elements is preferentially destroyed. Material that enters the ISM gas phase, either by direct transport from stellar environments or by the destruction of dust, is homogenised and anomalous isotopic signatures are lost. Therefore, incomplete condensation of elements around AGB stars and potential destruction of more labile stardust in the ISM can explain the smaller nucleosynthetic Pd variations relative to the more refractory Zr, Mo and Ru (Figure 2) and the lack of nucleosynthetic variations in more volatile elements such as Cd and Te.

**Disconnect between light and heavy nuclides**

Nucleosynthetic deficits smaller than those in Zr, Mo and Ru are reported for a number of heavier elements, e.g., Ba, Nd and W, while no variations are detected in Pt and Os[4,50]. Many of these elements are highly refractory and therefore these observations cannot wholly reflect incomplete condensation around AGB stars. For example, the condensation temperatures of ZrC, MoC, HfC and WC at C/O ratios typical of low mass AGB stars are all within the same range[48]. Instead, we propose that this trend was determined by the initial metallicity of the AGB stars that contributed the majority of stardust to the protoplanetary disk. The production of elements heavier than, and including, Ba ($Z \geq 56$; a magic neutron number) in AGB stars decreases as the metallicity increases, relative to Zr, Mo, Ru and Pd (Figure 4). This is a well-known feature of the $^{13}$C neutron source in AGB stars[51] (Extended Data Figure 5), because Fe seeds are more abundant in higher metallicity stars and capture more neutrons, which reduces the production of heavier s-process elements. Meanwhile, the production of SiC and silicate dust increases as a function of metallicity[52,53]. Hence, most *s*-process stardust in the solar protoplanetary disk likely originated from AGB stars with high metallicities, which was also suggested to account for the isotopic composition of SiC grains[54]. Models predicting the contribution of silicate and SiC stardust from AGB stars with different metallicities to the Solar System also support this hypothesis[37]. As a result, the concentrations of heavier elements ($Z \geq 56$) in stardust, relative to Zr, Mo and Ru, are lower than in the bulk Solar System *s*-process component. The heavier elements are predominantly carried by ISM dust that incorporated homogenised material from AGB stars with a wide range of metallicities. This leads to the smaller or non-detectable nucleosynthetic offsets in the heavier ($Z \geq 56$) refractory elements.

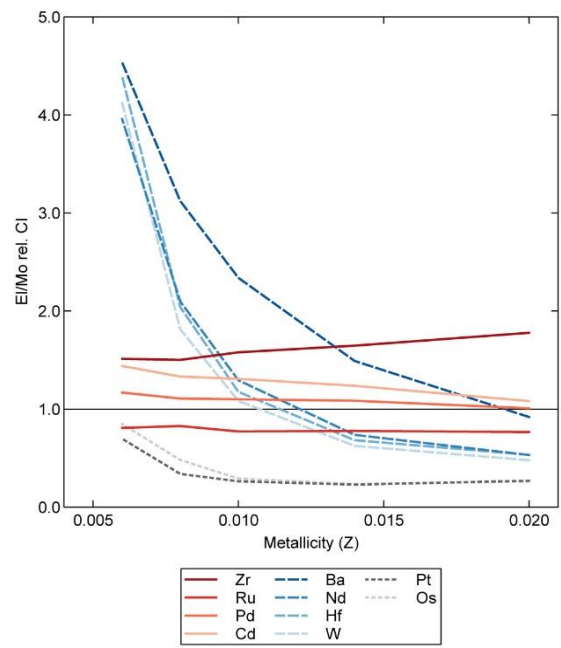

**Figure 4 | The *s*-process yield of elements relative to Mo and normalised to CI chondrites for a 3 M$_\odot$ AGB star with varying initial metallicities.** The *s*-process yield of the elements Zr (96 %*), Ru (37 %*), Pd (53 %*) and Cd (70 %*), relative to Mo (58 %*), are largely unaffected by the initial metallicity of AGB stars. The *s*-process yields of heavier elements Ba (89 %*), Nd (57 %*), Hf (61 %*), W (65 %*), Os (12 %*) and Pt (7 %*) decreases as the initial metallicity increases. The AGB stellar yields were taken from the FRUITY database[57] for non-rotating stars with an initial mass of 3 solar masses, CI values from Ref. 58. These trends are independent of the initial mass and are present in all current AGB datasets (see Extended Data Figure 5).* Estimated *s*-process contribution to the solar abundance from Ref. 20.

**Origin of the NC-CC dichotomy**

An extension to our model can account for the isotopic dichotomy noted between CC and NC meteorites[32], whereby CC groups are enriched in supernova derived material. A small pervasive enrichment of supernova derived material in the CC reservoir (Figure 3) recreates the enrichment of light neutron-rich isotopes (e.g., $^{50}$Ti or $^{54}$Cr) as well as the negative shift in $\varepsilon^{92}$Mo without changing the slope of the $\varepsilon^{92}$Mo-$\varepsilon^{100}$Mo correlation (Extended Data Figure 4) in agreement with meteorite data[11]. This can be attributed to a compositional change in the material infalling to the protoplanetary disk with time and/or by thermal processing[30,31], with complete homogenisation between the two reservoirs (CC and NC) prohibited by the formation of Jupiter's core[55,56]. Therefore, a change in the composition of the infalling material is a viable addition to the model presented here (Figure 3).

**Conclusions**

Our study represents a novel attempt to combine the nucleosynthetic variations identified in planetary bodies with data for presolar grains, results from astronomical observations on the origin and evolution of dust, and state-of-the-art nucleosynthetic models, to explain the origin of planetary isotopic heterogeneity in our Solar System. We present the first model that can simultaneously explain the origin of two prevalent nucleosynthetic features observed in rocky bodies: the subdued isotope variations in (i) more volatile elements and (ii) heavier ($Z \geq 56$) elements.

**Acknowledgements** This work was supported by the European Research Council under the European Union's Seventh Framework Programme (FP7/2007–2013)/ERC Grant agreement no. [279779] awarded to M.S. and by the Lendület grant (LP17-2014) of the Hungarian Academy of Sciences awarded to M.L. The authors acknowledge funding from ETH, the National Center for Competence in Research "PlanetS", supported by the Swiss National Science Foundation (SNSF), and project funding from the SNSF (200020_179129). We are grateful to D. Farsky and D. Cook for their assistance in acquiring the Rh/Pd ratios used in this study. We thank C. Smith and D. Cassey (Natural History Museum, London), J. Hoskin (Smithsonian Institute) and P. Heck (Field Museum) for the loan of meteorite materials used in this study. Comments from M. Rehkämper helped improve an early version of this manuscript.

**METHODS**

**Samples.** A total of 24 meteorites from the IAB, IIAB, IID, IIIAB, IVA and IVB iron meteorite classes were selected for Pd isotope analyses (Extended Data Figure 1; Supplementary Table 1). Platinum isotope data were collected from the same sample aliquots and are published in Ref. 50,59,60, except new Pt isotope data for the IVB iron meteorites Tlacotepec and Hoba, the IVA iron meteorite Muonionalusta and the IID iron meteorite Rodeo (Supplementary Table 1). Four samples from the IID iron meteorite Carbo were analysed to evaluate the effects of cosmic ray exposure (CRE). These samples were taken from a cross section through Carbo and the CRE magnitude of each sample, inferred from offsets in $\varepsilon^{182}W/^{184}W$ and $^3He$ concentration, increases in the order Carbo J > Carbo G > Carbo Y > Carbo A. Sampling locations are adjacent to those of Ref. 61.

**Chemical separation and isotope measurement by MC-ICP-MS.** Palladium and Pt were separated from a single sample aliquot using the methods described in Ref. 60,62. All Pd and Pt isotope measurements were performed at ETH Zürich using a Thermo Scientific Neptune *Plus* multi collector-inductively coupled plasma mass spectrometer (MC-ICPMS) coupled with a Cetac Aridus II desolvating introduction system and standard H cones. Results are reported in epsilon notation (ε), i.e. the deviation of the sample isotope ratio from the average of the bracketing NIST SRM 3138 (Pd) or NIST SRM 3140 (Pt) standards, given in parts per 10 000 relative to $^{105}$Pd or $^{195}$Pt. Palladium isotope determinations follow Ref. 62. Samples were bracketed by the NIST SRM 3138 Pd standard matched to within 15 % of the sample concentration. Sample solutions were diluted to ~100 ng ml$^{-1}$ Pd to achieve an ion beam intensity between 5 and 7 × 10$^{-11}$ A for $^{105}$Pd. All ratios were internally normalised to $^{108}$Pd/$^{105}$Pd = 1.18899 (Ref. 63) using the exponential law. Additional normalisations are available in Supplementary Table 1. Platinum isotope analyses followed the method outlined by Ref. 60. Data were corrected for instrumental mass bias using the exponential law, and were internally normalized to $^{198}$Pt/$^{195}$Pt = 0.2145 (Ref. 64). Sample solutions were diluted to give ion beam intensities of between ~2 x 10$^{-10}$ and 4 x 10$^{-10}$ A for $^{194}$Pt.

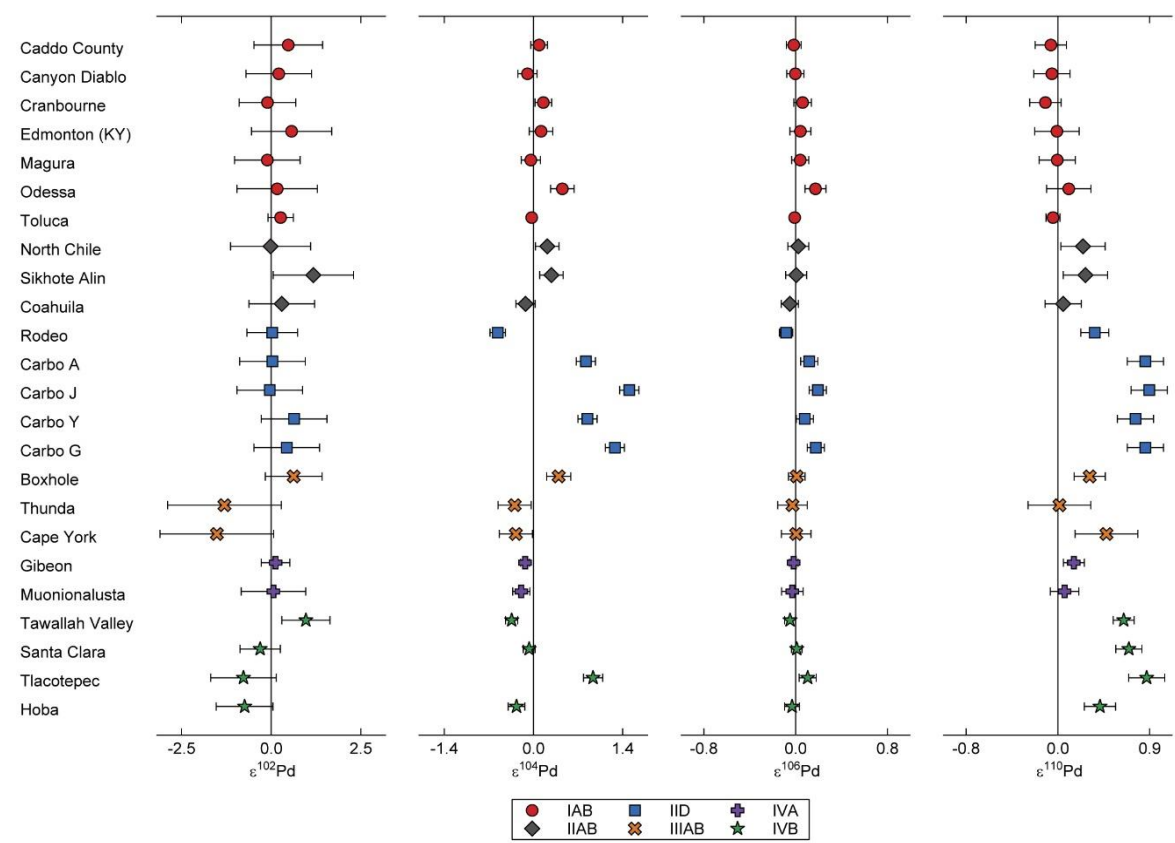

**Extended Data Fig. 1 | Palladium isotope composition of iron meteorites from the IAB, IIAB, IID, IIIAB, IVA and IVB groups.** All epsilon values are reported relative to $^{105}Pd$ and internally normalised to $^{108}Pd/^{105}Pd$. Uncertainties on data points reflects the 2 standard error of the mean.

**The reproducibility of the Pd isotope data.** The external reproducibility of Pd was estimated via a homoscedastic approach using the equations outlined in Ref. 65. By combining all samples with 4 or more analyses, we derived a general 2 standard deviation (*2 SD$_{external}$*) of 1.58 for $\varepsilon^{102}$Pd, 0.26 for $\varepsilon^{104}$Pd, 0.13 for $\varepsilon^{106}$Pd and 0.27 $\varepsilon^{110}$Pd. The uncertainty for each sample is reported as the 2 standard error of the mean (*2 SE$_{sample}$*). For samples where the number of analyses (*n*) is less than 4 the *2 SE$_{sample}$* is calculated using the *2 SD$_{external}$* (Equation 1). For samples where $n \geq 4$ the *2 SE$_{sample}$* is calculated using the 2 standard deviation of the individual analyses of the sample (*2 SD$_{sample}$*) or, if larger, *2 SD$_{external}$* (Equation 2).

$$n < 4: \quad 2\,SE_{sample} = 2\,SD_{external}/\sqrt{n_{sample}} \qquad (1)$$

$$n \geq 4: \quad 2\,SE_{sample} = \max(2\,SD_{sample}, 2\,SD_{external})/\sqrt{n_{sample}} \qquad (2)$$

**The reproducibility of the Pt isotope data.** The 2 standard deviation external precision was determined using repeat analyses of the IIAB iron meteorite North Chile (our in-house standard). Five aliquots of North Chile, passed through column chemistry independently, yield a precision of 0.49 for $\varepsilon^{192}$Pt, 0.17 for $\varepsilon^{194}$Pt and 0.07 for $\varepsilon^{196}$Pt (n=9)[50] for ion beam intensities of 4 x 10$^{-10}$ A.

**Elemental ratio measurements.** Several samples were analysed for their Rh/Pd ratios based on the method of Ref. 66. An aliquot of ~ 20 mg of material was taken from the digested sample prior to the chemical separation of Pd and Pt for isotope analysis. This was dried and then taken up in 0.5 ml 2 M HCl and refluxed on the hotplate at 100 °C overnight. Samples were then diluted to 2.5 ml with ultra-pure H$_2$O (MQ) and again refluxed. They were further diluted to 5 ml with MQ just prior to the ion exchange procedure. In preparation for the cation exchange, Bio-Rad™ 2 ml Poly-Prep® columns were rinsed with MQ and then loaded with 2 ml of Bio-rad™ AG® 50W-X8 200-400 mesh resin. The columns and resin were then rinsed with 5 ml MQ, 10 ml 4 M HCl and finally with another 10 ml MQ. The resin was pre-conditioned with 4 ml 0.2 M HCl, before loading the sample in 5 ml 0.2 M HCl and washing with 4 ml 0.2 ml HCl. These two fractions were combined and used to determine Rh/Pd ratios. After cation exchange chemistry, all samples were dried and re-dissolved twice with 0.5 ml 5 M HNO$_3$, before being diluted for analysis by a Thermo Scientific Element XR. Analyses were calibrated against synthetic standards and repeat analyses of two aliquots of the meteorite Odessa indicate a precision for the Rh/Pd ratio of better than 8 % (2 RSD). Blanks for this procedure are less than 1 ng for Rh and Pd, and hence negligible. The determined Rh/Pd ratios are within the range of those previously reported values for iron meteorites[67-69].

**Evaluating nuclear field shift effects.** Due to differences in the nuclear size and shape of isotopes, particularly between odd and even mass isotopes, mass-independent fractionation can occur during chemical exchange reactions[70]. Nuclear field shift (NFS) effects can vary for each aliquot that is processed through ion exchange chemistry, resulting in varying offsets for meteorites from the same group. Additionally, multiple aliquots of the same sample may also show variable Pd isotope compositions. The relative offset for each isotope can be modelled based on the nuclear charge radii using the equation presented in Ref. 71. The largest effect in Pd is expected on $^{105}$Pd, the sole odd isotope (Extended Data Figure 2). Such effects propagate on to all Pd isotope ratios when $^{105}$Pd is used for internal normalisation, and result in positive shifts in $\varepsilon^{102}$Pd, $\varepsilon^{104}$Pd, and $\varepsilon^{106}$Pd and a negative shift in $\varepsilon^{110}$Pd (Extended Data Figure 2a). A more suitable normalisation for detecting NFS effects is $^{108}$Pd/$^{106}$Pd where a large negative offset in $\varepsilon^{105}$Pd is expected (Extended Data Figure 2). The IAB irons show the smallest nucleosynthetic offsets of all the iron groups in other elements including, e.g. Mo and Ru[11,15], and therefore they are the ideal samples to look for NFS effects that arise during chemical separation. With one exception, all IAB samples, including the five individually processed individual aliquots of Toluca, are within uncertainty of the bracketing standard (Extended Data Figure 2b). Odessa is clearly resolvable from the other IAB meteorites, however, it does not fall on the predicted NFS trend suggesting the offset is not caused by such effects (Extended Data Figure 2). Similarly, the isotopic heterogeneities in the IID group between Rodeo and the four samples of Carbo (IID) are not consistent with the NFS trend (Extended Data Figure 2). These within-group variations are well matched by cosmic ray exposure effects. We therefore find no evidence for NFS effects on Pd isotopes in our data, and can exclude this process when interpreting variations in the iron meteorites.

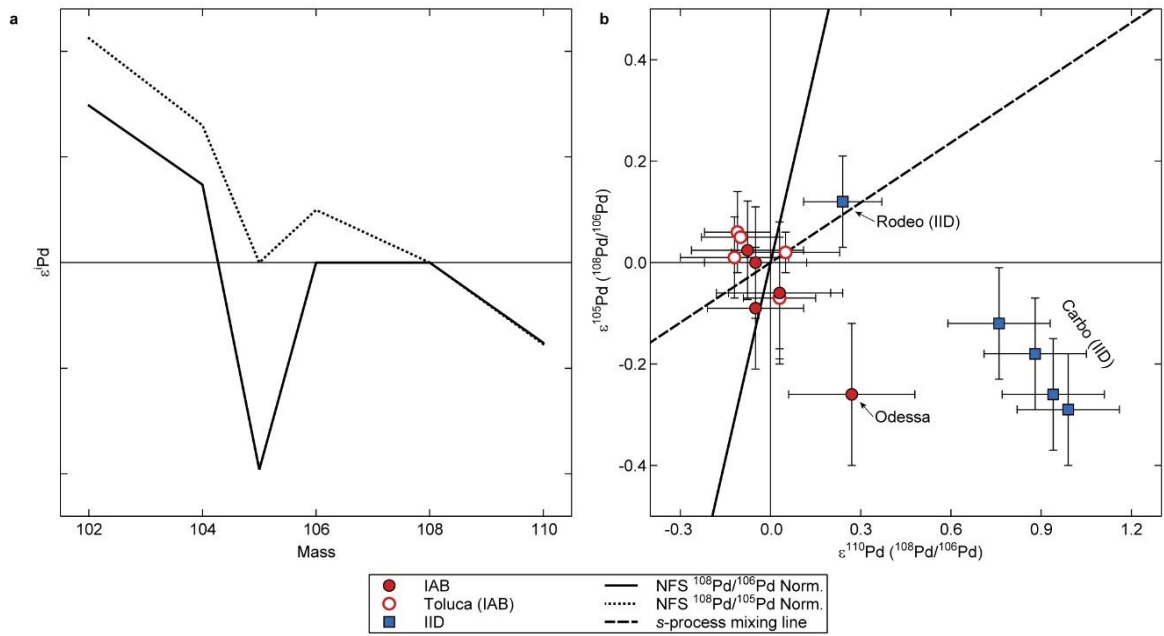

**Extended Data Fig. 2 | Nuclear field shift effects on Pd isotopes.** (**a**) The Pd isotope pattern produced by nuclear field shift effects, internally normalised to $^{108}$Pd/$^{105}$Pd (short dashed line) and $^{108}$Pd/$^{106}$Pd (solid line), calculated using the equations from Ref. 89 and the charge radii from Ref. 90. (**b**) $\varepsilon^{105}$Pd against $\varepsilon^{110}$Pd (internally normalised to $^{108}$Pd/$^{106}$Pd; Supplementary Table 1) for five individually processed aliquots of Toluca (IAB), Odessa (IAB), the other IAB meteorites, Rodeo (IID) and four aliquots of Carbo (IID) sampled at different locations within the meteorite. The solid line shows the nuclear field shift trend, internally normalised to $^{108}$Pd/$^{106}$Pd, and the dashed line shows an *s*-process deficit/excess trend calculated using the *s*-process yields of Ref. 20. Uncertainties are shown as the 2 standard error of the mean.

**Evaluating cosmic ray exposure effects.** Mass-independent isotope variation can occur from exposure to galactic cosmic rays as a meteoroid travels through space. Cosmic ray exposure (CRE) leads to the production of secondary neutrons upon interaction with a meteoroid[72]. Within-group variations for iron meteorites due to CRE are reported for a range of different elements, e.g., Mo, Ru, Pd, W, Os and Pt[13,19,50,61,64,73,74]. Here we use $\varepsilon^{196}$Pt as a neutron dosimeter to quantify the CRE effects on Pd isotopes. For Pd, the reaction $^{103}$Rh(n,β)$^{104}$Pd is important because of the large difference in relative abundance of $^{103}$Rh and $^{104}$Pd (100 % vs. 11 %), particularly as both elements occur at similar concentrations in iron meteorites[67,68]. Hence, Rh/Pd ratios are correlated with the CRE-induced effects on $\varepsilon^{104}$Pd. Modelling predicts that production from $^{103}$Rh is relatively large (3.5 $\varepsilon^{104}$Pd per 1 $\varepsilon^{196}$Pt when Rh/Pd = 1), while the production of $^{104}$Pd from Pd isotopes is low (0.1 $\varepsilon^{104}$Pd per 1 $\varepsilon^{196}$Pt)[72].

All Pt and Pd isotope data in this study were determined on the same sample aliquot and thus they can be compared without ambiguity. The CRE effects in each group were assessed by fitting a linear regression for each Pd isotope ratio against $\varepsilon^{196}$Pt (Extended Data Figure 3). The slope of the regression represents the $\varepsilon^{i}$Pd offset for every 1 $\varepsilon^{196}$Pt offset and the y-axis intercept defines the nucleosynthetic Pd isotope composition. Regressions for the IAB, IID and IVB groups all yield well-defined slopes (Extended Data Figure 3), while the slopes for the IIAB, IIIAB, and IVA irons show very large uncertainties because of the narrow spread of $\varepsilon^{196}$Pt values (Δ $\varepsilon^{196}$Pt < 0.3; Supplementary Table 2), i.e. all samples within these groups had similar exposure to CRE. No variations in $\varepsilon^{102}$Pd were identified and this suggests that $\varepsilon^{102}$Pd is not affected by CRE at our current level of precision, which is supported by CRE modelling[72] (Extended Data Figure 3). For $\varepsilon^{104}$Pd vs. $\varepsilon^{196}$Pt, the regression must include the variable Rh/Pd ratios of the samples[72]. Five samples do not have measured Rh/Pd ratios (see Supplementary Table 1). Instead, a value was estimated based on the average of other samples in the same group. With the exception of Carbo A and Carbo J, all samples without Rh/Pd ratios have low $\varepsilon^{196}$Pt values and therefore their Rh/Pd has little influence on the slopes calculated. The Rh/Pd ratio was measured for Carbo Y and Carbo G and these give similar Rh/Pd ratios.

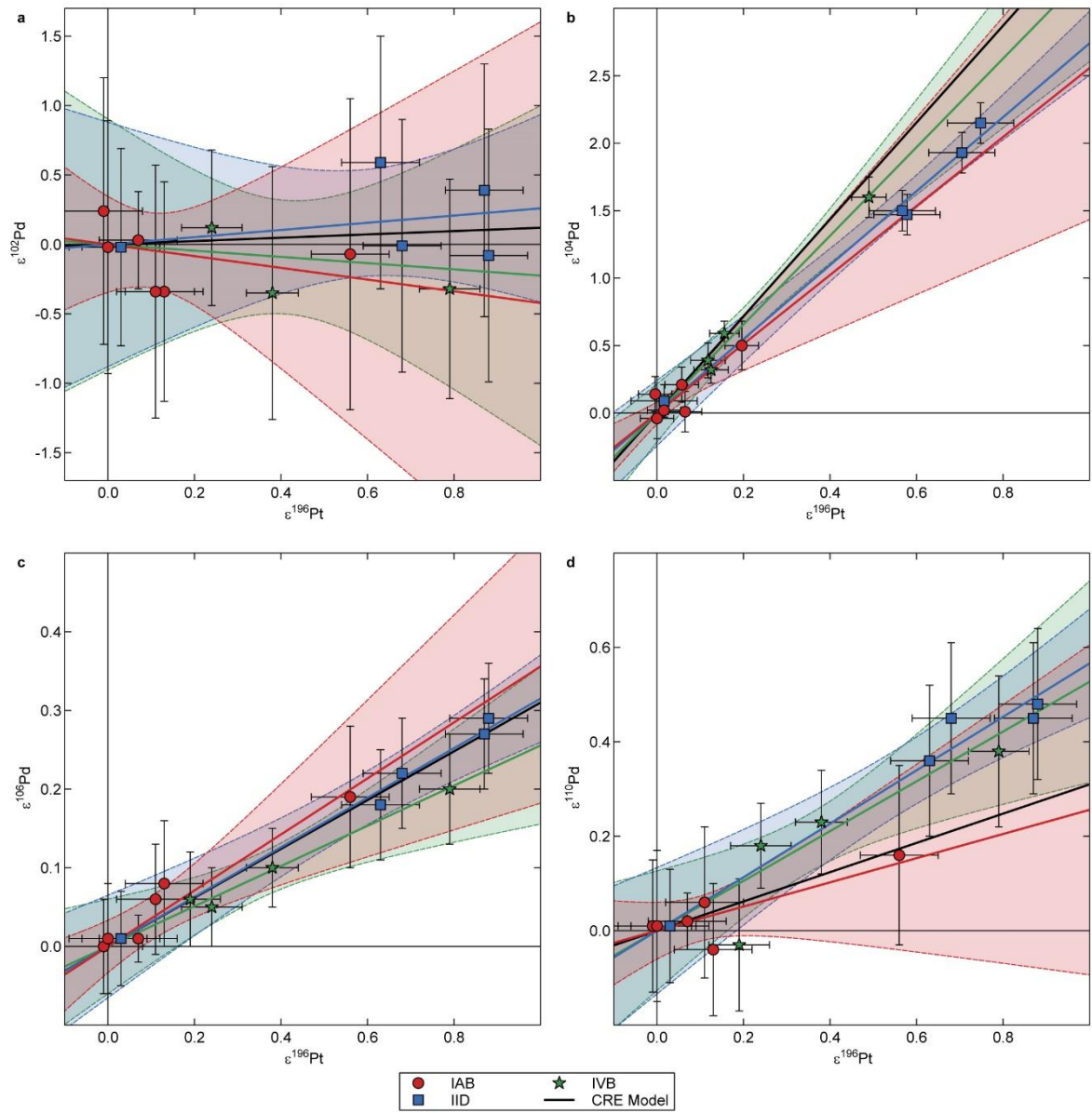

**Extended Data Fig. 3 | Cosmic ray effects (CRE) on Pd isotopes in iron meteorites.** Regressions of (**a**) $\varepsilon^{102}Pd$, (**c**) $\varepsilon^{106}Pd$, and (**d**) $\varepsilon^{110}Pd$ against $\varepsilon^{196}Pt$ for the IAB, IID and IVB groups. Panel (**b**) shows regressions of $\varepsilon^{104}Pd$ versus $\varepsilon^{196}Pt$ multiplied by the Rh/Pd ratio of the sample, to account for varied CRE contributions from $^{103}Rh(n,\beta)^{104}Pd$, for the same three groups. Individual samples and the slope of the regression are normalised to the intercept for each isotope/group such that the slopes can be compared directly. The black line shows the modelled CRE trend for each isotope, taken from Ref. 72. The slope of the regressions for $\varepsilon^{102}Pd$, $\varepsilon^{104}Pd$, $\varepsilon^{106}Pd$ and $\varepsilon^{110}Pd$ overlap for all three groups and agree well with the modelled slope. Uncertainties on individual data points given as the 2 standard error of the mean. Uncertainty envelope around regressions represents the 2 standard deviations of the regression calculated using the equation from Ref. 91.

Given the close sampling locations and similar $\varepsilon^{104}$Pd of Carbos Y and G and Carbos A and J, the average Rh/Pd values are suitable for the latter samples. The $\varepsilon^{196}$Pt value of each sample ($\varepsilon^{196}Pt_{sample}$) was then normalised to the median Rh/Pd ratio ($Rh/Pd_{median}$) of all samples belonging to the same group, using the following equation:

$$\varepsilon^{196}Pt_{normalised} = \varepsilon^{196}Pt_{sample} \times \frac{(Rh/Pd)_{sample}}{(Rh/Pd)_{median}} \qquad (3)$$

where $\varepsilon^{196}Pt_{normalised}$ is the renormalized $\varepsilon^{196}$Pt value of the sample and $Rh/Pd_{sample}$ is the Rh/Pd ratio of the sample. After applying this formula, the best-fit line for each group was obtained and the resulting slopes were re-normalised to Rh/Pd = 1 for ease of comparison (Supplementary Table 2). The slopes of the $\varepsilon^{104}$Pd - $\varepsilon^{196}$Pt correlations for the IAB, IID and the IVB irons are within uncertainty of each other (~2.6 ± 1.3 for Rh/Pd = 1; Extended Data Figure 3b). The regression for $\varepsilon^{106}$Pd against $\varepsilon^{196}$Pt yields consistent slopes across all three groups of ~0.30 ± 0.15 (Supplementary Table 2; Extended Data Figure 3c). The IID and IVB irons define slopes of ~0.55 ± 0.30 in the $\varepsilon^{110}$Pd - $\varepsilon^{196}$Pt diagram (Extended Data Figure 3d), while the IABs have a shallower slope of 0.26 ± 0.38. However, all three slopes are within uncertainty of each other (Supplementary Table 2; Extended Data Figure 3d). Overall, the consistency of the $\varepsilon^{104}$Pd, $\varepsilon^{106}$Pd and $\varepsilon^{110}$Pd variations against $\varepsilon^{196}$Pt for the different groups strongly supports CRE as the source of within-group Pd isotope variations. These variations also agree well with the modelled CRE effects by Ref. 72, providing further support of this conclusion (Supplementary Table 2; Extended Data Figure 3).

**Correcting for cosmic ray effects.** To determine the nucleosynthetic isotope composition of a group requires either samples with no CRE contribution, or alternatively a method for correction of CRE effects. In reality, all samples may have experienced some exposure to CRE and a tolerance limit for what can be considered negligible relative to the analytical uncertainties was defined. For this study the tolerance limits, i.e. the maximum $\varepsilon^{196}$Pt value where a sample can be considered unexposed, are 0.03 for $\varepsilon^{104}$Pd (Rh/Pd = 1), 0.08 for $\varepsilon^{106}$Pd and 0.1 for $\varepsilon^{110}$Pd. These numbers were empirically estimated based on the typical $\varepsilon^{196}$Pt value when the Pd isotope composition of a sample falls outside the uncertainty of the regression intercept, i.e. the unexposed Pd isotope composition (Supplementary Table 2). The lack of covariation between $\varepsilon^{102}$Pd and $\varepsilon^{196}$Pt implies that all $\varepsilon^{102}$Pd data in this study can be considered unexposed. Unexposed values are preferentially selected over calculated intercepts to determine the nucleosynthetic composition of each group, when available. When more than one sample falls below the $\varepsilon^{196}$Pt limits stated above a weighted mean is used to calculate the unexposed composition. All groups, with the exception of the IVB, have at least one sample with a $\varepsilon^{196}$Pt value that falls below the limits stated above. All groups apart from the IAB irons show nucleosynthetic effects that are distinct from the isotopic composition of the terrestrial standard in the order IVB ≥ IID > IIIAB > IVA > IIAB ≥ IAB = Earth (Figure 1; Table 1).

**Calculating dust removal.** The cosmic abundances of elements heavier than Fe are produced by three types of processes: the *s*-process, the *r*-process, and the *p*-process. Each of these is a collection of several processes occurring in different stellar environments. Traditionally, three *s*-process components appeared to be present in the Solar System. The production of the elements from Fe to Sr has been attributed to the weak component, which occurs in the core He- and shell C-burning phases of massive stars via the activation of the $^{22}$Ne neutron source[75]. The abundances of the heavier elements have been attributed to the main and strong components (the latter referring specifically to Pb), which are well known to occur in low mass AGB stars (~ 1.5 to 4 M$_\odot$) through the activation of the $^{13}$C neutron source[76]. The $^{22}$Ne neutron source is also the dominant neutron source in more massive AGB stars (> 4 M$_\odot$; Ref. 77,78), however, these stars do not produce enough *s*-process nuclei to be major contributors to the elements heavier than Fe in the Galaxy[35,79-81]. Early galactic chemical evolution models of the neutron-capture elements have shown that roughly 30% of the solar elemental abundance of the first s-process peak elements (Sr, Y, and Zr) is not accounted for and proposed the Light Element Primary Process (LEPP) to produce the missing abundances[35]. Proposed sources for the LEPP do not necessarily produce typical s-process *isotopic* signatures and are usually related to massive and/or low-metallicity objects in the early Universe, because it was originally noticed that such LEPP components may be identical to those observed in very old stars in the halo of our Galaxy. However, the need for this extra component is still debated[82]. Another possible contribution to the *s*-process was identified in low-metallicity, massive stars that rotate much faster than massive stars in the local Universe (Spinstars[36]). This allows them to produce *s*-process elements beyond Sr and contribute to the main component of the Solar System composition[83-86]. For the *r*-process, a weak (up to Ba) and a strong component are invoked to explain observations of *elemental* abundances in metal-poor stars. Both components could have multiple stellar origins and various *isotopic* signatures, different from the residual Solar System *r*-process isotopic composition (from core-collapse supernovae neutrino driven winds[87], to electron capture supernovae[88]). For the strong component, neutron star mergers now appear to be an observationally confirmed site[28,29]. The *p*-process also has likely more than one component and stellar site, ranging from the γ-disintegration process in thermonuclear and core-collapse supernovae, to neutrino winds and the alpha-rich freeze-out in core-collapse supernovae (see e.g. Ref. 27).

In spite of this complexity, to which we will return below, as a first approximation let us consider a simplified scenario whereby the mass (*x*) of a planetary body with an average Solar System composition ($x_{solarbody}$) can be summarised as the combination of the three traditional components:

$$x_{solarbody} = x_s + x_r + x_p \qquad (4)$$

where $x_s$, $x_r$ and $x_p$ is the mass contributed by the different *s*-, *r*-, and *p*-processes discussed above, respectively. Nucleosynthetic variations in the Solar System are the result of a heterogeneous distribution of one or more of these three components, or their subcomponents.

The nucleosynthetic variations observed in rocky bodies and meteorites can be described as a relative excess, or deficit, of the solar *s*-process component (e.g. Ref. 4). Using $x_{solarbody}$ as our starting point we can describe the mass of this body ($x_{rockybody}$), but with a given nucleosynthetic variation as:

$$x_{rockybody^1} = (1+z)x_s + x_r + x_p \tag{5}$$

where $z$ is the excess/deficit of *s*-process material as a fraction of Solar System *s*-process component (I.e. $x_s$ in Eq. 4). In the model presented here (see main text) we propose that nucleosynthetic offsets are the result of a relative enrichment of stardust. This can be written as:

$$x_{rockybody^1} = (1+y)x_{stardust} + x_{ISM\ dust} \tag{6}$$

where $x_{stardust}$ and $x_{ISM\ dust}$ is the mass of stardust and ISM dust, respectively, and $y$ is the excess/deficit of stardust as a fraction of Solar System stardust component. Assuming that stardust is entirely comprised of *s*-process material, $x_{ISM\ dust}$ be described as:

$$x_{ISM\ dust} = (x_s - x_{stardust}) + x_r + x_p \tag{7}$$

We can then solve for $y$ by combining Eq. 5, Eq. 6 and Eq. 7:

$$y = z\frac{x_s}{x_{stardust}} \tag{8}$$

Nucleosynthetic variations are typically attributed to the removal of material in the solar nebula rather than the addition of material. To achieve this we rewrite Eq. 5 and Eq. 6 as:

$$x_{rockybody^2} = x_s + \frac{x_r + x_p}{(1+z)} \tag{9}$$

$$x_{rockybody^3} = x_{stardust} + \frac{x_{ISM\ dust}}{(1+y)} \tag{10}$$

Note that while the relative mass of the rocky bodies in Eq. 5 and Eq. 6 are identical, this is not true for Eq. 9 and Eq. 10. However, the relative proportion of the different nucleosynthetic components remains the same for the rocky body produced all 4 equations. Because $x_{rockybody}$ is the same as $x_{solarbody}$ when $z = 0$ and $y = 0$ we can calculate the mass fraction of dust that has been removed ($f_{lost}$) to create a planetary body with a given nucleosynthetic composition using the following equation:

$$f_{lost} = 1 - \frac{x_{rockybody}}{x_{solarbody}} \tag{11}$$

The largest uncertainty in these calculations is the assumption that the stardust material being removed/enriched has the same composition as the bulk $x_s$, $x_r$ or $x_p$ components. The composition of stardust is difficult to constrain as there are few data are available in literature. For example, if the abundance of trace elements in stardust/ISM dust is only half that of the bulk Solar System composition, then $f_{lost}$ increases by a factor of 2 in both scenarios discussed in the main text. The presence of small amounts of *p*- and *r*-process stardust would also lower the magnitude of the overall *s*-process composition of stardust, again increasing the value of $f_{lost}$.

**Author Contributions** M.S. designed the research project. M.E. prepared the samples for isotope analyses and conducted the measurements with assistance from A.H. M.E. did the data interpretation and wrote the first draft of the manuscript with important input from M.S., A.H. and M.L. All authors contributed equally to subsequent revisions of the manuscript.

**Author Information** Correspondence and requests for materials should be addressed to Mattias.ek@bristol.ac.uk

**Data availability.** Original Pd and Pt data points for individual meteorites is also available from the EarthChem library (https://doi.org/doi:10.1594/IEDA/111416). All other data are available from the corresponding author upon reasonable request.

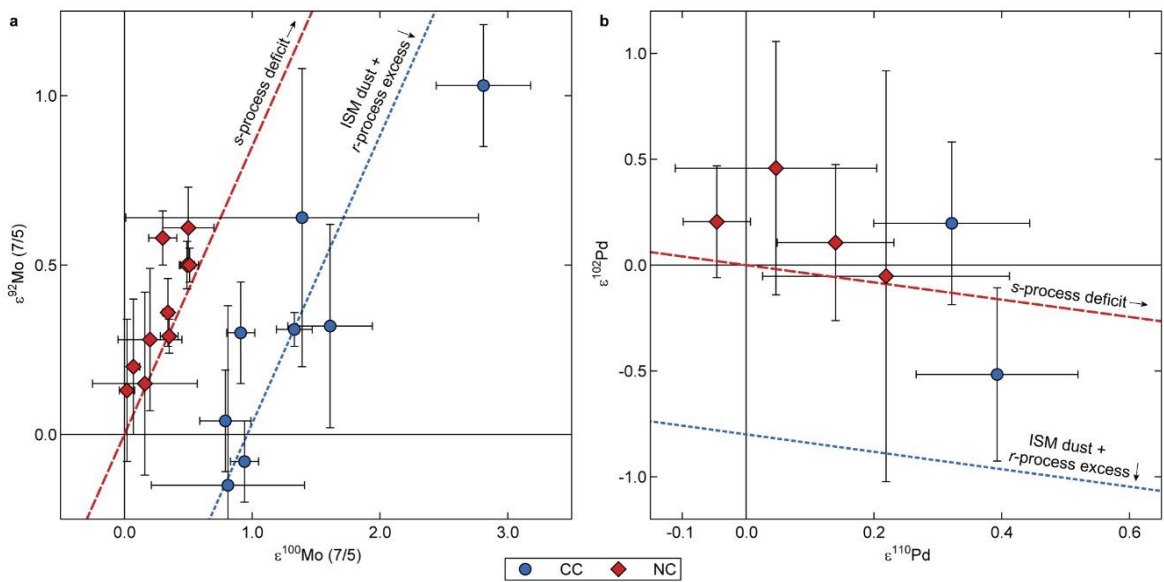

**Extended Data Fig. 4 | Isotopic dichotomy between carbonaceous (CC) and non-carbonaceous (NC) meteorites in ε$^{100}$Mo-ε$^{92}$Mo and ε$^{110}$Pd-ε$^{102}$Pd.** (**a**) The dichotomy reported in Mo is characterised by an enrichment in ε$^{92}$Mo for the CC meteorites (blue) relative to the NC group (grey). A small addition of supernova derived material to the stardust and/or ISM dust fraction coupled with thermal processing of ISM dust mantles can explain this offset. (**b**) Only the IVB irons of the two analysed CC-type iron meteorite groups (IID and IVB) show the negative shift in ε$^{102}$Pd predicted by the isotopic dichotomy. Given the typical uncertainty on ε$^{102}$Pd for individual meteorites (~ 1 ε; Supplementary Table 1) due to the large Ru correction on $^{102}$Pd (Ref. 62), it is barely possible to resolve the expected effect. The dashed lines indicate a mixing line between an *s*-process endmember[33] and the terrestrial composition. The blue dashed line represents a mixing line between an *s*-process endmember[33] and the terrestrial composition with a 0.008% enrichment in the residual *r*-process component, estimated based on the Mo data. Mo data from Ref. 11,14 and Pd data from Table 1. Uncertainties on Pd data points reflect either the 2 standard error of the mean or the 2 standard deviation of the *x*-axis intercept of a regression against ε$^{196}$Pt (See Table 1). Uncertainties on Mo data points reflect either the 2 standard error of the mean (data from Ref. [11]) or the 95 % confidence interval (data from Ref. 14).

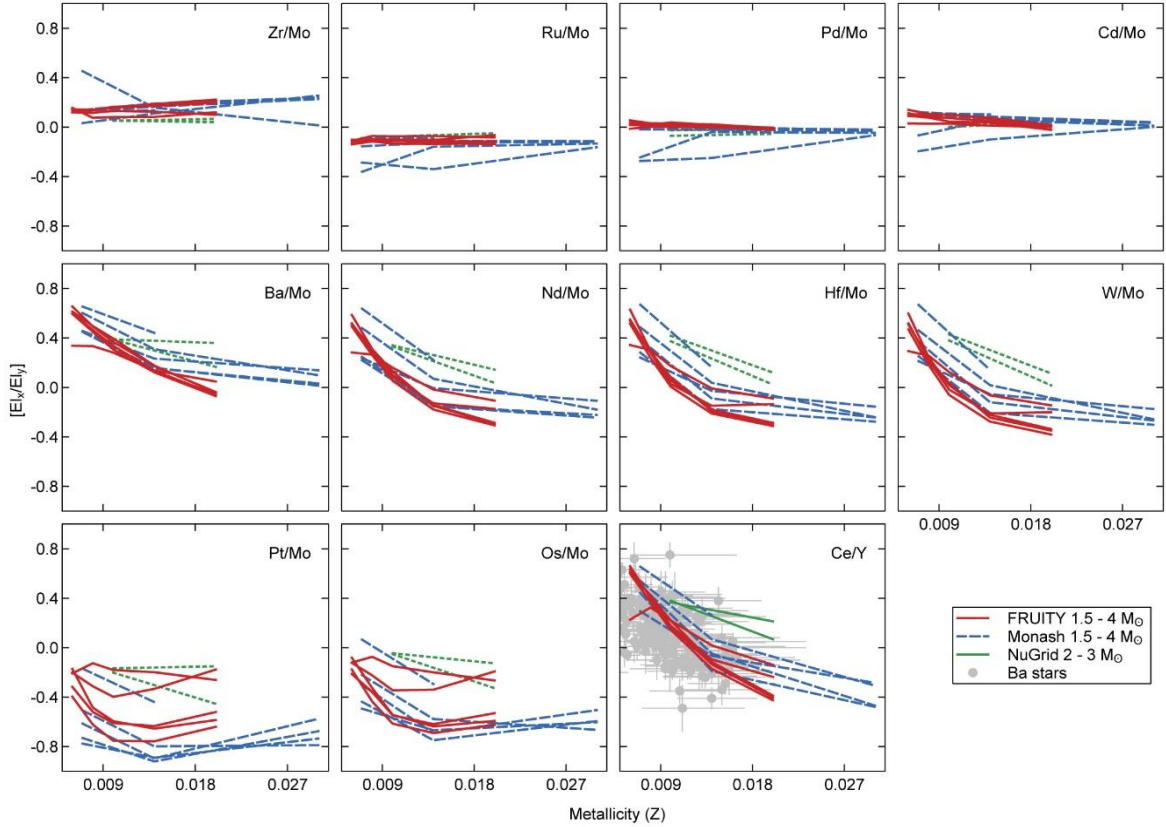

**Extended Data Fig. 5 | Elemental ratios as a function of metallicity for FRUITY, Monash and NuGrid *s*-process models for AGB stars with an initial mass between 1.5 – 4 M$_\odot$.** The relative proportion of light *s*-process elements (Y, Zr, Mo, Ru, Pd and Cd) vary little with different metallicities and are independent of the initial stellar mass and nucleosynthetic model. All models show a clear trend with the yield of heavy *s*-process elements (Ba, Ce, Nd, Hf, W, Pt, Os) decreasing, relative to the light *s*-process elements, as the metallicity increases. Shown in panel Ce/Y are the observational data for Ba stars[51,92] in grey. These also indicate a decrease in the Ce/Y ratio as a function of increasing metallicity[51]. FRUITY data[93-95] correspond to the total yield for non-rotating stars with a metallicity of 0.006, 0.08, 0.010, 0.014 and 0.020 and mass of 1.5, 2.0, 2.5, 3.0 and 4.0 M$_\odot$. Monash data[81,96-98] depict the yields of stars with a metallicity of 0.007, 0.014 and 0.030 and mass of 1.5, 2.5, 3.0, 3.5 and 4.0 M$_\odot$ computed with a mass extension of the mixing leading to the formation of the main neutron source $^{13}$C of $2\times10^{-3}$ (M$_\odot \leq 3$) and $1\times10^{-3}$ (M$_\odot > 3$). NuGrid data[99] represent the final surface composition for stars with a metallicity of 0.01 and 0.02 and mass of 2.0 and 3.0 M$_\odot$. Elemental ratios are shown in standard spectroscopic notation where [El$_x$/El$_y$]=log(El$_x$/El$_y$)$_*$-log(El$_x$/El$_y$)$_\odot$, where El$_x$ and El$_y$ are abundances by number.